\documentclass[aps,prb,twocolumn,groupedaddress,showpacs]{revtex4}

\usepackage{graphicx}
\usepackage{dcolumn}
\usepackage{bm}

\begin{document}

\title{Microscopic analysis of multipole susceptibility of actinide dioxides:
\\ A scenario of multipole ordering in AmO$_2$}

\author{Takashi Hotta}

\affiliation{Department of Physics, Tokyo Metropolitan University,
Hachioji, Tokyo 192-0397, Japan}

\date{\today}

\begin{abstract}
By evaluating multipole susceptibility of
a seven-orbital impurity Anderson model
with the use of a numerical renormalization group method,
we discuss possible multipole states of actinide dioxides
at low temperatures.
In particular, here we point out a possible scenario
for multipole ordering in americium dioxide.
For Am$^{4+}$ ion with five $5f$ electrons, it is considered
that the ground state is $\Gamma_7^{-}$ doublet and
the first excited state is $\Gamma_8^{-}$ quartet,
but we remark that the $f^5$ ground state is easily converted
due to the competition between spin-orbit coupling and Coulomb interactions.
Then, we find that the $\Gamma_8^-$ quartet can be the ground state of AmO$_2$
even for the same crystalline electric field potential.
In the case of $\Gamma_8^-$ quartet ground state,
the numerical results suggest that high-order multipoles
such as quadrupole and octupole can be relevant to AmO$_2$.
\end{abstract}



\maketitle

\section{Introduction}

Actinide dioxides with the fluorite structure of
the space group $Fm3m$ have been studied intensively
for more than fifty years
both from experimental and theoretical sides.\cite{review1,review2}
A typical target material is UO$_2$
mainly due to its technological importance
as a nuclear reactor fuel and a heterogeneous catalyst.
As for theoretical research, a clear picture of
the electronic structure in UO$_2$ has been obtained.
In fact, neutron scattering results show that UO$_{2}$ is
a noncollinear antiferromagnet below 30.8K.\cite{Caciuffo1}
Detailed analysis of core photoemission spectra has
suggested that UO$_{2}$ is an insulator
of Mott-Hubbard type.\cite{Roy}
The crystalline electric field (CEF) states of UO$_2$
have been also determined.\cite{Amoretti}

Concerning NpO$_2$, over fifty years since 1953,\cite{Westrum}
it has been known to exhibit a mysterious low-temperature
ordered phase.\cite{Ross,Erdos}
Several phenomenological works on the ordered phase have
claimed a key role of octupole degree of freedom.
\cite{Santini,Paixao,Caciuffo2,Lovesey,Kiss}
The CEF states of NpO$_2$ have been determined
by neutron scattering experiment.\cite{Fournier}
Recently, the octupole order has been strongly supported
by $^{17}$O-NMR experiment \cite{Tokunaga}
and by inelastic neutron scattering study.\cite{Magnani}
In order to understand why such high-order multipole ordering appears,
it is necessary to proceed to the research in a microscopic level.
For this issue, it has been shown that octupole order actually
characterizes the ground state of NpO$_2$ by the analysis of
an $f$-electron model on the basis of a $j$-$j$ coupling scheme
on an fcc lattice.\cite{Kubo1,Kubo2}

On the other hand, PuO$_2$ is known to be a semiconductor
with magnetic susceptibility
which is almost independent of temperature up to 1000 K,
since the CEF ground state is $\Gamma_1^+$ singlet and
the first excited state is $\Gamma_4^+$ triplet
with the large excitation energy as 123 meV.\cite{Kern1,Kern2}
Thus, from the viewpoint of magnetism,
PuO$_2$ did not attract much attention.

Now let us turn our attention to AmO$_2$.
Probably due to the difficulty in the treatment of this material
with high radioactivity,
we cannot find lots of experimental results on AmO$_2$.
In 1969, M\"ossbauer isomer shift in AmO$_2$ was measured.\cite{Kalvius}
After that, the magnetic susceptibility was measured and the peak was
found around at 15 K.\cite{Karraker}
Naively thinking, such a peak seems to suggest the signal of
antiferromagnetic ordering,
while neutron diffraction measurement did not detect antiferromagnetic
order in agreement with the M\"ossbauer measurement.\cite{Boeuf}
This situation looks similar to that of NpO$_2$.
Namely, multipole degree of freedom seems to be
a key issue in AmO$_2$ to reconcile experimental results,
as has been proposed for NpO$_2$ in the context of octupole ordering.

Here we note that the CEF ground state of AmO$_2$ was considered
to be $\Gamma_7^-$ doublet
from the experimental results,\cite{Karraker,Abraham,Kolbe}
but it could not bring higher multipoles.
This is in sharp contrast to the case of NpO$_2$ with
the confirmed CEF ground state of $\Gamma_8^-$ quartet.
Due to the CEF analysis of actinide dioxides,
the $\Gamma_7^-$ doublet ground state has been suggested.
\cite{Magnani2}
Thus, it seems to be the mainstream in the research of actinide
dioxides to clarify a mechanism which explains
the disappearance of antiferromagnetic order
for the $\Gamma_7^-$ doublet ground state.

However, we believe that there still exists an alternative scenario
on the basis of multipole ordering in AmO$_2$,
when we recall the fact that the CEF ground state
of the $f^5$ electron system
is easily converted due to the competition between
spin-orbit coupling and Coulomb interaction.\cite{Hotta1}
This point has been also discussed by the present author
to propose a possible scenario which explains
the change of the CEF ground state among Sm-based filled
skutterudite compounds,\cite{Hotta2}
since trivalent Sm ion includes five $4f$ electrons.
It has been experimentally \cite{Moore1,Moore2,Moore3,Moore4}
and theoretically \cite{Moore1,Moore2,Moore3,Moore4,Shim}
shown that the angular momentum coupling of the $5f$ states of
Am is situated between the $LS$ and $j$-$j$ coupling limits
for many chemical situations, albeit closer to the $j$-$j$ limit.
Thus, a strong spin-orbit coupling is present in the $5f$ states
of $f^5$ and $f^6$ configurations of Am that can compete with
Coulomb interactions.
Accordingly, it is believed to be meaningful to pursue
a possibility of multipole ordering
in AmO$_2$ with the $\Gamma_8^-$ quartet ground state.

In this paper, we show that multipole order is possible in
AmO$_2$, when we appropriately take into account
both spin-orbit coupling and Coulomb interaction
in the $f$-electron terms for the CEF ground state.
In the case of five $f$ electrons such as Am$^{4+}$ ion,
the ground state is easily converted between $\Gamma_7^-$ and $\Gamma_8^-$,
when spin-orbit coupling and Coulomb interaction compete with each other.
Then, the $\Gamma_8^-$ quartet can be the ground state,
even if the CEF potential is unchanged.
In order to see what type of multipole is relevant,
we evaluate the multipole susceptibility of
the Anderson model by using a numerical renormalization group method.
We find that higher-order multipoles are actually relevant to AmO$_2$
within the present calculation.

The organization of this paper is as follows.
In Sec.~II, we discuss the local $f$-electron state emerging from
the competition among the Coulomb interaction and spin-orbit
coupling under the CEF potential.
In particular, the change of the $f^5$ ground state is
explained in detail.
In Sec.~III, we show the model Hamiltonian.
In order to discuss multipole properties,
it is necessary to define the multipole operator.
Here we explain the description of the multipole
as spin-charge complex one electron operator.
Then, we briefly explain the numerical technique
used in this paper.
In Sec.~IV, we show the results of the multipole state
for the case of $n=2 \sim 5$, where $n$ is the local $f$-electron number.
In particular, the results for $n$=5 are discussed in detail.
Finally, the paper is summarized in Sec.~V.
Throughout this paper,
we use such units as $\hbar$=$k_{\rm B}$=1.

\section{Local $f$-electron state}

Let us first discuss the local $f$-electron states of actinide ions.
The local Hamiltonian should be composed of three parts as
\begin{equation}
  H_{\rm loc}=H_{\rm so}+H_{\rm int}+H_{\rm CEF}.
\end{equation}
The first term denotes the spin-orbit coupling, given by
\begin{eqnarray}
  H_{\rm so} =
  \lambda \sum_{m,m'}\sum_{\sigma,\sigma'}
  \zeta_{m,\sigma,m',\sigma'}
  f_{m\sigma}^{\dag}f_{m'\sigma'},
\end{eqnarray}
where $\sigma$=$+1$ ($-1$) for up (down) spin,
$f_{m\sigma}$ is the annihilation operator for $f$ electron with
spin $\sigma$ and $z$-component $m$ of angular momentum $\ell$=3,
$\lambda$ is the spin-orbit coupling,
$\zeta_{m,\pm 1,m,\pm 1}=\pm m/2$,
$\zeta_{m \pm 1,\mp 1,m, \pm 1}=\sqrt{12-m(m \pm 1)}/2$,
and zero for the other cases.

The second term indicates the Coulomb interaction among
$f$ electrons, expressed as
\begin{eqnarray}
  H_{\rm int} =
  \sum_{m_1 \sim m_4}\sum_{\sigma,\sigma'}
  I_{m_1m_2,m_3m_4}
  f_{m_1\sigma}^{\dag}f_{m_2\sigma'}^{\dag}
  f_{m_3\sigma'}f_{m_4\sigma},
\end{eqnarray}
where the Coulomb integral $I_{m_1,m_2,m_3,m_4}$ is given by
\begin{eqnarray}
  I_{m_1,m_2,m_3,m_4}=\sum_{k=0}^{6} F^k c_k(m_1,m_4)c_k(m_2,m_3).
\end{eqnarray}
Here $F^k$ is the radial integral for the $k$-th partial wave,
called Slater integral or Slater-Condon parameter \cite{Slater,Condon}
and $c_k$ is the Gaunt coefficient.\cite{Gaunt,Racah}
Note that the sum is limited by the Wigner-Eckart theorem to even values
($k$=0, 2, 4, and 6).

The third term is the CEF potential,
given in the one-electron potential form as
\begin{eqnarray}
  H_{\rm CEF} =
  \sum_{m,m'}\sum_{\sigma}
  B_{m,m'}f_{m\sigma}^{\dag}f_{m'\sigma},
\end{eqnarray}
where $B_{m,m'}$ is the CEF potential.
Since the fluorite structure belongs to $O_{\rm h}$ point group,
$B_{m,m'}$ is given by using a couple of CEF parameters
$B_4^0$ and $B_6^0$ for angular momentum $\ell$=3 as \cite{Hutchings}
\begin{eqnarray}
  \begin{array}{l}
    B_{3,3}=B_{-3,-3}=180B_4^0+180B_6^0, \\
    B_{2,2}=B_{-2,-2}=-420B_4^0-1080B_6^0, \\
    B_{1,1}=B_{-1,-1}=60B_4^0+2700B_6^0, \\
    B_{0,0}=360B_4^0-3600B_6^0, \\
    B_{3,-1}=B_{-3,1}=60\sqrt{15}(B_4^0-21B_6^0),\\
    B_{2,-2}=300B_4^0+7560B_6^0.
  \end{array}
\end{eqnarray}
Note the relation of $B_{m,m'}$=$B_{m',m}$.
Following the traditional notation,\cite{LLW} we define
\begin{eqnarray}
  \label{CEFparam}
  B_4^0=Wx/F(4),~B_6^0=W(1-|x|)/F(6),
\end{eqnarray}
where $x$ and the sign of $W$ specify the CEF energy scheme,
while $|W|$ determines the energy scale for the CEF potential.
Concerning non-dimensional parameters, $F(4)$ and $F(6)$,
we use $F(4)$=15 and $F(6)$=180 for $\ell$=3.\cite{Hutchings}

Here we briefly explain the parameters of the local Hamiltonian.
Concerning Slater-Condon parameters,
first we set $F^0$=10 eV by hand,
since we are not interested in the determination of
the absolute value of the ground state energy.
Others are determined so as to reproduce excitation spectra of
U$^{4+}$ ion with two $5f$ electrons.\cite{Eliav}
Here we show only the results:
$F^2$=6.36 eV, $F^4$=5.63 eV, and $F^6$=4.13 eV.
As for spin-orbit coupling $\lambda$,
we use the values of actinide atoms
such as $\lambda$=0.235 eV (U), 0.272 eV (Np),
0.311 eV (Pu), and 0.351 eV (Am).

For the estimation of the CEF parameters,
let us summarize the CEF energy levels of actinide dioxides.
For UO$_2$, the ground state is $\Gamma_5^+$ triplet and
the first excited state is $\Gamma_3^+$ doublet with the
excitation energy 150 meV.\cite{Amoretti}
For NpO$_2$, the ground and first excited states are,
respectively, $\Gamma_8^{-(2)}$ and $\Gamma_8^{-(1)}$
quartets with the excitation energy 55 meV.\cite{Fournier}
For PuO$_2$, the ground state is $\Gamma_1^+$ singlet,
while the first excited state is $\Gamma_4^+$ triplet
with the excitation energy 123 meV.\cite{Kern1,Kern2}

Now we set the CEF parameters for actinide dioxides
in the present notation.
For the purpose, first we estimate $W$ and $x$
so as to reproduce the CEF scheme of UO$_2$ by
the $f^2$ electrons state.
This is not a difficult task,
since the CEF parameters are easily restricted from
the experimental results.
After that, among the values of $W$ and $x$
appropriate for UO$_2$,
we further restrict the values of $W$ and $x$
which can also reproduce the results for NpO$_2$ and PuO$_2$.
Note that the CEF states for $f^3$ and $f^4$ states are
almost reproduced by using the parameters of $f^2$ electron state.
Since the CEF term is just given by the one-electron potential,
the CEF effect is not so drastically changed among
the materials with the same crystal structure,
even though the $f$-electron number is different.

After some calculations,
we determine $W$=$-10.5$ meV and $x$=0.62.
The results for the CEF level schemes are summarized in Fig.~1.
It is observed that the CEF states of UO$_2$, NpO$_2$,
and PuO$_2$ are well reproduced.
Note here that for the same CEF parameters,
the ground state for Am$^{4+}$ is found to be
$\Gamma_7^-$ doublet
and the excited state is $\Gamma_8^-$ quartet
with the excitation energy of about 50 meV.
This is consistent with the previous theoretical results
on the CEF states of AmO$_2$
obtained by more detailed calculations.\cite{Magnani2}

\begin{figure}[t]
\includegraphics[width=8.5cm]{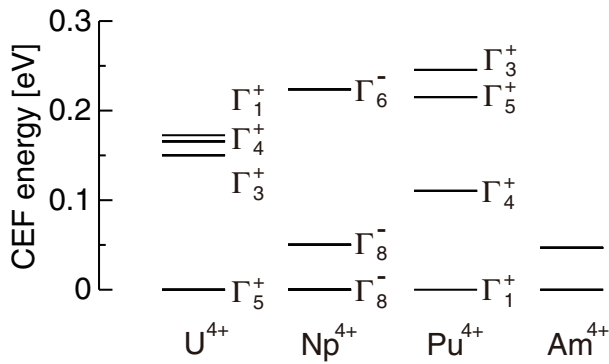}
\caption{
CEF energy level schemes for tetravalent actinide ions.
Parameters used here are explained in the maintext.
}
\end{figure}

Naively thinking, there occurs
ordering of magnetic moment originating from
$\Gamma_7^-$ ground state for AmO$_2$,
but it seems to be contradict with neutron diffraction study.
In order to resolve such contradiction,
there are two ways:
One is to consider a mechanism which explains the
disappearance of magnetic moment even in the
$\Gamma_7^-$ ground state.
Another way is to reconsider the local $f$-electron term
by focusing on spin-orbit coupling
and Coulomb interaction.
Here we propose an alternative scenario
on the basis of the second direction.

Thus far, we have simply assumed that Coulomb interaction
is not changed among different actinide ions,
but in actuality, they may be changed.
Concerning spin-orbit coupling, we have
used the values in actinide atom, but
it may be also changed in the ionic states.
Then, we point out that for $f^5$ systems,
the CEF states are sensitively changed
by the competition between Coulomb interaction and
spin-orbit coupling in comparison with other values of
local $f$-electron number.\cite{Hotta1,Hotta2}

In order to understand that
the effect of Coulomb interaction and
spin-orbit coupling appears in the CEF parameters,
here we express the CEF parameter $B_4^0$ by using the
so-called Stevens factor as
\begin{equation}
   B_4^0= A_4 \langle r^4 \rangle \beta^{(n)}_{J},~
\end{equation}
where $A_k$ is the parameter depending on materials,
$\langle r^k \rangle$ denotes the radial average of
local $f$-electron wavefunction,
$n$ denotes the local $f$-electron number,
$J$ is the total angular momentum of the ground state multiplet,
and $\beta^{(n)}_{J}$ indicates the Stevens factor,
which is one of coefficients
appearing in the method of Stevens' operator equivalent.\cite{Stevens}

For the case of $n$=5,
it is well known that the ground state multiplet is
characterized by $J$=5/2.
After lengthy calculations, for $n$=5 and $J$=5/2,
we can obtain that
$\beta^{(5)}_{5/2}$=$(13/21)\beta_{3}$
in the $LS$ coupling scheme and
$\beta^{(5)}_{5/2}$=$-(11/7)\beta_{3}$
in the $j$-$j$ coupling scheme,
\cite{Hotta1}
where $\beta_{3}$ denotes the Stevens factor
for $\ell$=3, given by $\beta_{3}$=$2/495$.
It should be noted that the sign of $\beta^{(n)}_{J}$ is
changed between the $LS$ and $j$-$j$ coupling schemes,
suggesting that the ground state is converted,
when Coulomb interaction and/or spin-orbit coupling are changed.

The modification of Coulomb interaction and spin-orbit coupling
is closely related to the picture for multi-$f$-electron state.
The CEF level schemes in Fig.~1 is qualitatively
understood by a $j$-$j$ coupling scheme.
Namely, by assuming that the effective Hund's rule coupling
is smaller than the CEF level splitting,
we simply accommodate plural numbers of $f$ electrons
in the levels of $\Gamma_8^-$ ground
and $\Gamma_7^-$  excited states.
Then, we can easily reproduce all the CEF level schemes
of tetravalent actinide ions.
However, the conversion of the CEF ground state for $n$=5
indicates that the actual situation should be slightly
shifted to the side of the $LS$ coupling scheme.
We note that due to such a shift,
the CEF states for $n$=2, 3, and 4
are not qualitatively changed, while in the case of
$n$=5, the ground state is converted.

\begin{figure}[t]
\includegraphics[width=8.5cm]{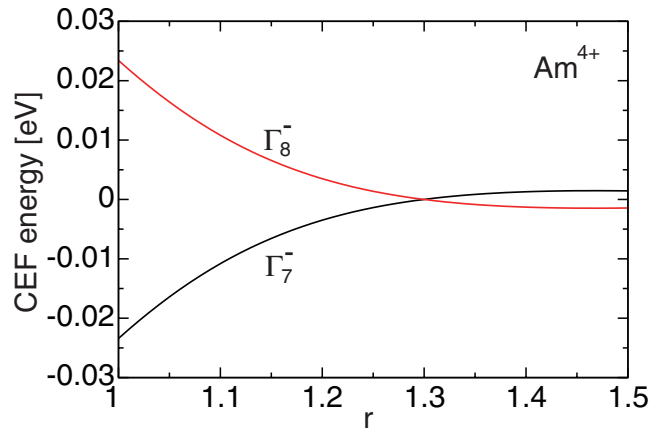}
\caption{
The CEF energy for $n$=5 vs. $r$.
The meaning of $r$ is explained in the maintext.
}
\end{figure}

Here we emphasize that such a ground-state conversion occurs
in the region of realistic values of Coulomb interaction and
spin-orbit coupling,
as has been pointed out in the discussion of the
CEF states of Sm-based filled skutterudites.\cite{Hotta2}
In fact, as mentioned in Sec.~I, it has been shown that
the angular momentum coupling of the $5f$ states of Am
is situated between the $LS$ and $j$-$j$ coupling limits,
but it is rather closer to the $j$-$j$ coupling limit.
\cite{Moore1,Moore2,Moore3,Moore4,Shim}
%
Thus, it seems to be reasonable to change slightly
spin-orbit coupling and/or Coulomb interaction
in the case of tetravalent Am ion.

If we resort to first-principles calculation,
we may determine correctly $F^j$ and $\lambda$,
but it is out of the scope of the present paper.
Here we simply introduce an artificial parameter
$r$ to control Coulomb interactions and spin-orbit coupling as
$F^k \rightarrow r F^k$ and $\lambda \rightarrow \lambda/r$
with $r \ge 1$.

In Fig.~2, we depict the energies of $H_{\rm loc}$
as functions of $r$ for $n$=5.
We find that the conversion of the ground state
occurs around at $r \approx 1.3$.
Note that in other values of $n$, the ground states
are not changed.
Thus, we can obtain the $\Gamma_8^-$ quartet
ground state for Am$^{4+}$.
In the following sections, we will discuss
how the multipole states actually appear
when we change the value of $r$ for Am ion.

\section{Model and Method}

\subsection{Anderson model}

Now we include the hybridization between localized and
conduction electrons.
The Hamiltonian is the Anderson model, given by
\begin{equation}
  \label{model}
  H \!=\! \sum_{{\bf k},\sigma}
  \varepsilon_{{\bf k}} c_{{\bf k}\sigma}^{\dag} c_{{\bf k}\sigma}
  +\sum_{{\bf k},\sigma,m}
  (V_{m} c_{{\bf k}\sigma}^{\dag}f_{m\sigma}+{\rm h.c.})
  +H_{\rm loc},
\end{equation}
where $\varepsilon_{\bf k}$ denotes conduction electron dispersion,
$c_{{\bf k}\sigma}$ indicates the annihilation operator for conduction
electron with momentum ${\bf k}$ and spin $\sigma$,
and $V_{m}$ is the hybridization between conduction and $f$ electrons.

Note that we consider only $a_{\rm u}$ single conduction band
with xyz symmetry composed of oxygen $2p$ electrons.
Since oxygen ions surrounding actinide ions are located
in the [1, 1, 1] direction, there should exist a conduction band
composed of $2p$ electrons with xyz symmetry.
This picture seems to be consistent with band-structure
calculation,\cite{Maehira}
but the ignorance of $t_{\rm 1u}$ and $t_{\rm 2u}$ bands is
just assumption.
Here we note that the hybridization occurs between the states
with the same symmetry of local $f$-electron state.
Since the $a_{\rm u}$ conduction band has xyz symmetry,
we set $V_2$=$-V_{-2}$=$V$ and zero for other $m$.
Hereafter, a half of the bandwidth of $a_{\rm u}$
conduction band $D$ is set as the energy unit,
i.e., $D$=1 eV.
We fix $V$ as $V/D$=0.05 throughout this paper.
Note that in order to adjust the local $f$-electron number $n$,
we appropriately change the chemical potential in the
actual calculation,
although we do not explicitly show such a term.

\subsection{Multipole operator}

In order to discuss multipole properties,
it is necessary to define the multipole operator
${\hat X}$ for $f$ electron.
\cite{Hotta3,Hotta4,Hotta5,Hotta6,Hotta7,Hotta8}
In general, ${\hat X}$ is expressed as
\begin{equation}
  \label{multi}
  {\hat X}=\sum_{k,\gamma}
   p^{(k)}_{\gamma}{\hat T}^{(k)}_{\gamma},
\end{equation} 
where $k$ is a rank of multipole,
$\gamma$ is a label to express $O_{\rm h}$ irreducible representation,
and ${\hat T}^{(k)}_{\gamma}$ is cubic tensor operator,
given by
${\hat T}^{(k)}_{\gamma}=\sum_q G^{(k)}_{\gamma,q}{\hat T}^{(k)}_{q}$.
Here an integer $q$ runs between $-k$ and $k$,
${\hat T}^{(k)}_q$ is spherical tensor operator,
and $G^{(k)}_{\gamma,q}$ is the transformation matrix
between spherical and cubic harmonics.
We determine $p^{(k)}_{\gamma}$ later.

In order to obtain explicit expression of the spherical tensor
operator ${\hat T}^{(k)}_q$,
it is convenient to convert the $f$-electron basis from $(m,\sigma)$
to $(j,\mu)$, where $j$ is the total angular momentum and
$\mu$ is the $z$-component of $j$.
When we define $f_{j\mu}$ as the annihilation operator
for $f$ electron labeled by $j$ and $\mu$,
we obtain ${\hat T}^{(k)}_q$ in the second-quantized form as
\begin{equation}
  {\hat T}^{(k)}_q = \sum_{j,\mu,\mu'}
  T^{(k,q)}_{j;\mu,\mu'}f^{\dag}_{j\mu}f_{j\mu'}.
\end{equation}
Note that there are no components between different values of $j$,
since the matrix for total angular momentum
is block-diagonalized in the basis of $(j,\mu)$.
The matrix element of $T^{(k,q)}_{j;\mu,\mu'}$ is calculated
by the Wigner-Eckart theorem as
\begin{equation}
  T^{(k,q)}_{j; \mu,\mu'}=
  \frac{\langle j || T^{(k)} || j \rangle}{\sqrt{2j+1}}
  \langle j \mu | j \mu' k q \rangle,
\end{equation}
where $\langle j \mu | j \mu' k q \rangle$ indicates
the Clebsch-Gordan coefficient
and $\langle j || T^{(k)} || j \rangle$ denotes
the reduced matrix element
for spherical tensor operator, given by
$\langle j || T^{(k)} || j \rangle$=
$\sqrt{(2j+k+1)!/(2j-k)!}/2^k$.
Note that $k$$\le$$2j$ and the highest rank of $f$-electron
multipole is 7.

Let us now determine the coefficient $p^{(k)}_{\gamma}$.
In order to discuss the multipole state, it is necessary to evaluate
the multipole susceptibility in the linear response theory.
However, multipoles belonging to the same symmetry are mixed in general,
even if the rank is different.
In addition, multipoles are also mixed due to the CEF effect.
Thus, we determine $p^{(k)}_{\gamma}$ by the normalized eigenstate
of susceptibility matrix
\begin{eqnarray}
  \label{sus}
  \chi_{k\gamma,k'\gamma'}
  \! &=& \! \frac{1}{Z}
  \sum_{i,j} \frac{e^{-E_i/T}-e^{-E_j/T}}{E_j-E_i}
  \langle i | [{\hat T}^{(k)}_{\gamma} \!-\! \rho^{(k)}_{\gamma}] | j \rangle
  \nonumber  \\ \! &\times& \!
  \langle j | [{\hat T}^{(k')}_{\gamma'} \!-\! \rho^{(k')}_{\gamma'}]| i \rangle,
\end{eqnarray}
where $E_i$ is the eigenenergy for the $i$-th eigenstate
$|i\rangle$ of $H$, $T$ is a temperature,
$\rho^{(k)}_{\gamma}$=$\sum_i e^{-E_i/T}
\langle i |{\hat T}^{(k)}_{\gamma}| i \rangle/Z$,
and $Z$ is the partition function given by
$Z$=$\sum_i e^{-E_i/T}$.
Note that the multipole susceptibility is given by
the eigenvalue of the susceptibility matrix.

\subsection{Method}

In order to evaluate multipole susceptibility of
the Anderson model, we employ
a numerical renormalization group (NRG) method.\cite{NRG}
in which momentum space is logarithmically discretized
to include efficiently the conduction electrons near the Fermi energy
and the conduction electron states
are characterized by ``shell'' labeled by $N$.
The shell of $N$=0 denotes an impurity site described by
the local Hamiltonian.

In the NRG method, we transform the Hamiltonian
into the recursion form as
\begin{eqnarray}
  H_{N+1} = \sqrt{\Lambda}H_N+t_N \sum_\sigma
  (c_{N\sigma}^{\dag}c_{N+1\sigma}+c_{N+1\sigma}^{\dag}c_{N\sigma}),
\end{eqnarray}
where $\Lambda$ denotes a parameter for logarithmic discretization,
$c_{N\sigma}$ indicates the annihilation operator of conduction electron
in the $N$-shell, and
$t_N$ is the hopping of electron between
$N$- and $(N+1)$-shells, expressed by
\begin{eqnarray}
  t_N=\frac{(1+\Lambda^{-1})(1-\Lambda^{-N-1})}
  {2\sqrt{(1-\Lambda^{-2N-1})(1-\Lambda^{-2N-3})}}.
\end{eqnarray}
The initial term $H_0$ is given by
\begin{eqnarray}
  H_0=\Lambda^{-1/2}[H_{\rm loc} + \sum_{\sigma}
  V(c_{0\sigma}^{\dag}f_{{\rm c}\sigma}
  +f_{{\rm c}\sigma}^{\dag}c_{0\sigma})].
\end{eqnarray}

Each component of multipole susceptibility matrix
eq.~(\ref{sus})
is evaluated by using the renormalized state.
Then, the multipole state is defined by the eigen states
of eq.~(\ref{sus}).
We note that the temperature $T$ is defined as
$T$=$\Lambda^{-(N-1)/2}$ in the NRG calculation,
where $N$ is the number of the renormalization step.
Due to the limitation of computer resources,
we keep only $M$ low-energy states.
In this paper, we set $\Lambda$=5 and $M$=3000.

\begin{figure}[t]
\includegraphics[width=8.5cm]{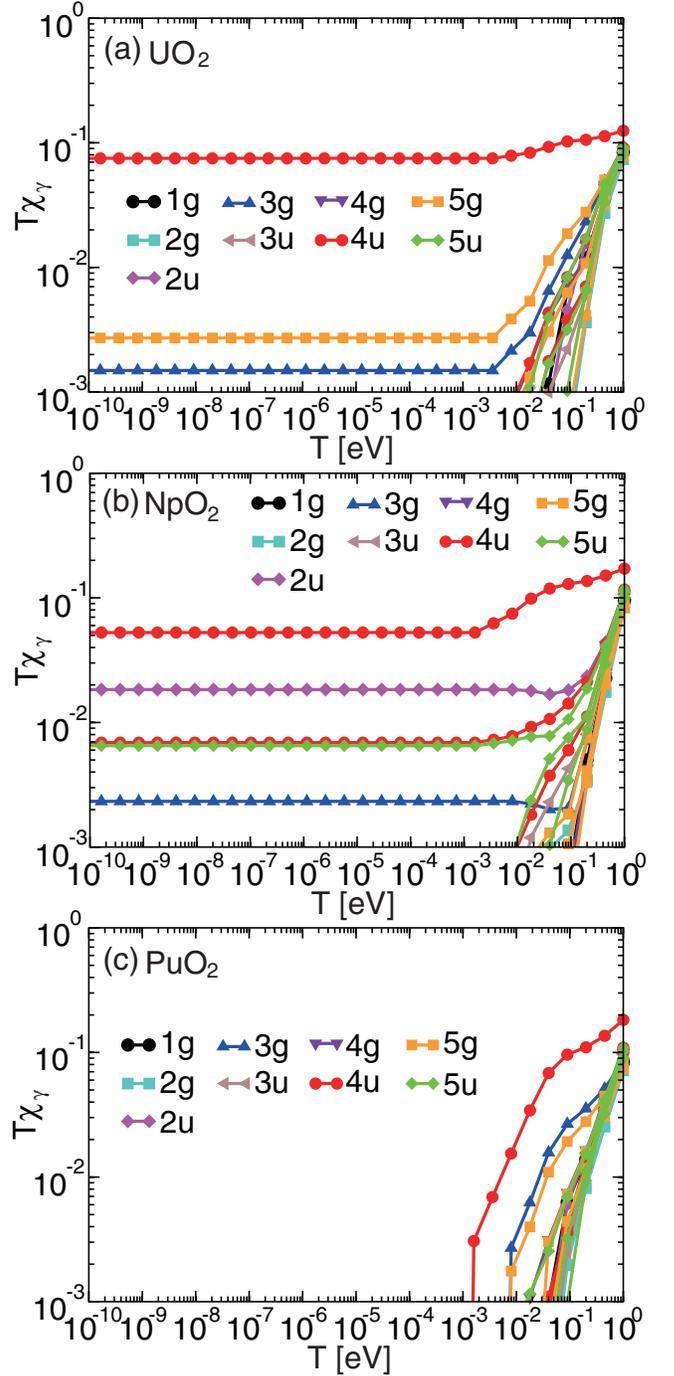}
\caption{$T\chi_{\gamma}$ vs. temperature $T$ for
(a) UO$_2$, (b) NpO$_2$, and (c) PuO$_2$.}
\end{figure}

\section{Results}

Now we discuss the multipole state of the Anderson model
eq.~(\ref{model}).
In Figs.~3(a)-(c), we show $T\chi_{\gamma}$ vs. $T$ for $n$=2, 3, and 4,
where $\chi_{\gamma}$ is the eigenvalue of the multipole susceptibility.
We use the same values of the parameters in the Hamiltonian
for $n$=2, 3, and 4.
The values of the spin-orbit coupling are
$\lambda$=0.235 eV, 0.272 eV, and 0.311 eV
for $n$=2, 3, and 4, respectively.
The eigenstates are classified by irreducible representation
of $O_{\rm h}$ point group.
Here we use short-hand notations such as ``3g'' and ``5u'',
which denote $\Gamma_3^+$ and $\Gamma_5^-$, respectively,
in the Bethe notation.
Note that ``1u'' does not appear among multipoles up to rank 7.

For $n$=2 (UO$_2$), we find the optimized state is labeled by 4u,
which is mainly composed of dipole (about $92 \%$).
The secondary components are quadrupoles (5g and 3g),
but $\chi_{\rm 5g}$ and $\chi_{\rm 3g}$ is smaller in one order
in comparison with $\chi_{\rm 4u}$.
For $n$=3 (NpO$_2$), as easily deduced from the $\Gamma_8^-$
quartet ground state, we find varieties of multipoles.
Among them, the primary component is 4u,
which is mainly composed of dipole (about $96 \%$).
The secondary one is 2u octupole.
In the third group, another 4u and 5u are almost degenerate.
We note that this 4u is composed of higher multipole component
and the 5u is mainly composed of octupole.
The fourth component is 3g quadrupole.
We emphasize the existence of octupoles (2u and 5u) with
significant eigenvalues.

In the present calculation, we cannot determine the kind
of multipole ordering in actual systems.
However, the multipoles which remain at low temperatures
are the candidates which will order
in the actual system.
In the case of NpO$_2$, it has been gradually revealed
that triple-${\bf q}$ order of 5u octupole can naturally
reconcile several kinds of experiments.
The 5u octupole is actually included in the multipoles
in the present calculations, although it is not dominant
component.

For $n$=4 (PuO$_2$), we do not find any significant multipole
component, as easily understood from the $\Gamma_1^+$ singlet
ground state which is well separated from the magnetic excited state.
In this sense, from the viewpoint of magnetism,
this case does not attract much attention.

\begin{figure}[t]
\includegraphics[width=8.5cm]{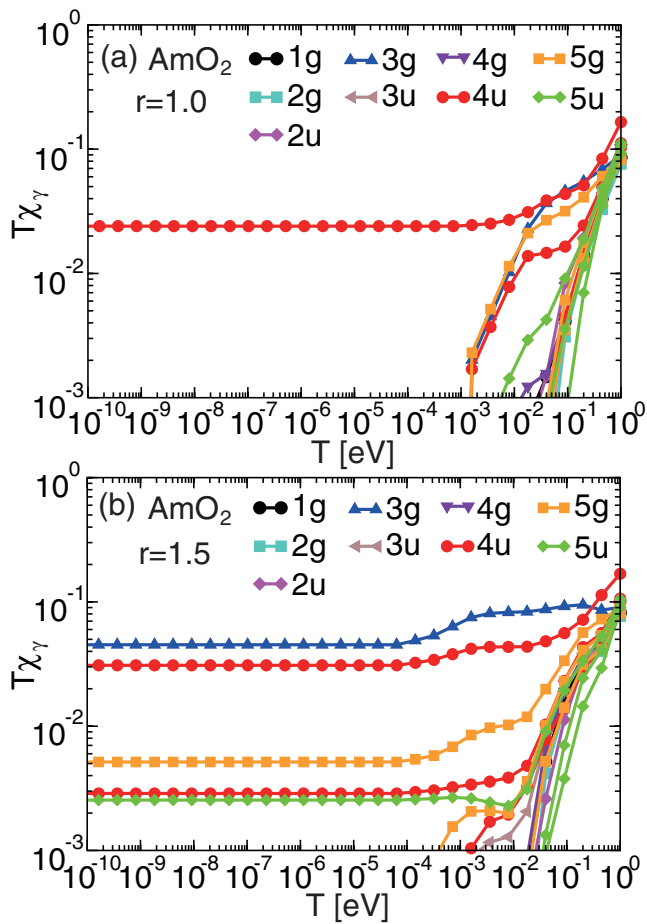}
\caption{(a) $T\chi_{\gamma}$ vs. temperature $T$ for $n$=5.
The parameters are the same as those in Fig.~2, except for
the spin-orbit coupling.
(b) $T\chi_{\gamma}$ vs. temperature $T$ for $n$=5 and
$r$=1.5 with enhanced Coulomb interaction and reduced
spin-orbit coupling.
}
\end{figure}

Next we move on to the case of $n$=5,
corresponding to AmO$_2$.
In Fig.~4(a), we show the results for $n$=5
by using the parameters for $r$=1.
Namely, the Coulomb interactions are the same
as those in Figs.~3(a)-(c).
The spin-orbit coupling is set as $\lambda$=0.351 eV,
which is the value for Am atom.
In this case, since the ground state is $\Gamma_7^-$ doublet,
the component which remain in the
low-temperature region is 4u, which is composed
of dipole (about 25$\%$) and octupole (about 75$\%$).
We note that the octupole component is significantly large
in comparison with the 4u states of UO$_2$ and NpO$_2$.
In any case, as we have expected,
we find only magnetic 4u moment
originating from $\Gamma_7^-$ ground state for AmO$_2$.

Then, we increase the value of $r$ in order to
move to the side of the $LS$ coupling scheme.
In Fig.~4(b), we plot $T\chi_{\gamma}$ of
the multipole susceptibility for $n$=5 and $r$=1.5
with the $\Gamma_8^-$ ground state (see Fig.~2).
In this case, we find that the primary component is 3g,
which is mainly composed of quadrupole.
The secondary component is 4u and
we also find 5g, 4u, and 5u components
with smaller eigenvalues.
We note that some multipoles of AmO$_2$ are
the same as those in NpO$_2$, except for 2u and 5g,
although the corresponding eigenvalues are different.

Now we provide a comment on the value of $r$,
which is introduced so as
to increase the effect of Coulomb interactions and
decrease the magnitude of spin-orbit coupling.
Note, however, that the angular momentum coupling of
the $5f$ states of Am is nowhere near as close to
the $LS$ coupling limit as for Cm.\cite{Moore3,Shim}
Thus, we should not entirely or strongly suppress
the spin-orbit coupling in the $5f$ states of Am.
It is acceptable that we change slightly
spin-orbit coupling and Coulomb interaction.
Accordingly, when the value of $r$ is increased
so as to move towards the side of the $LS$ coupling scheme,
we should pay due attention $not$
to suppress spin-orbit coupling too much.

\section{Discussion and Summary}

In this paper, we have discussed the multipole state in the
low-temperature region,
by analyzing the seven-orbital impurity Anderson model
with the use of the NRG method.
We have found the multipole state for $n$=2, 3, and 4,
which are not in contradiction to the phases observed in
UO$_2$, NpO$_2$, and PuO$_2$, respectively.
Note here that we determine the candidates which will
order in the actual periodic system at low temperatures.

For the case of $n$=5, when we use the same parameters
as those for $n=2 \sim 4$, we have suggested the phase
dominated by magnetic moment.
However, if we change slightly Coulomb interactions
and spin-orbit coupling, we have found the $\Gamma_8^-$
quartet ground state for AmO$_2$.
In this situation, we have shown that the low-temperature
phase can contain multipoles such as
quadrupole and octupole.
The parameters are artificially introduced here,
but our purpose is to point out a possibility of
the $\Gamma_8^-$ quartet ground state
due to the competition between Coulomb interactions and
spin-orbit coupling.

Unfortunately, we cannot determine the kind of multipole
order only from the present calculation,
but on the basis of the same crystal structure,
it is plausible that 5u octupole order also appears in
AmO$_2$.
On the other hand, it may be possible to exploit
other scenarios, e.g., quadrupole ordering,
which were invented for understanding of NpO$_2$.
In any cases,
the combination of phenomenological theory and
microscopic experiment will be useful to finalize
the kind of multipole which orders at low temperatures
in AmO$_2$.

Experimentally it has been considered that the CEF ground state
of AmO$_2$ is $\Gamma_7^-$ ground state.
However, we believe that it is still meaningful to examine
the experimental results on the basis of
the $\Gamma_8^-$ quartet ground state,
although it may be difficult to perform the microscopic
experiments of AmO$_2$.

Finally, let us provide a comment on the simplification
of the model.
In this paper, since we have considered only single conduction band,
there exists residual entropy in the results.
In actuality, it should be finally released when we
consider $t_{\rm 1u}$ and $t_{\rm 2u}$ conduction bands.
This point is also related to the relevant multipole
moment when we consider the ordered state
in the periodic systems.
In this sense, the present results are qualitative,
but they include the actual multipole
which forms ordered state.

In summary, we have discussed the multipole state of
actinide dioxides due to the evaluation of
the multipole susceptibility of the Anderson model.
When Coulomb interaction and spin-orbit coupling
have been appropriately changed,
it has been found that multipoles including quadrupole and
octupole are relevant to AmO$_2$.
It is believed that multipole ordering can be detected in
AmO$_2$ in future experiments.

\acknowledgments

The author is grateful to S. Kambe and Y. Tokunaga
for fruitful discussions and useful comments.
This work has been supported by a Grant-in-Aid for
for Scientific Research on Innovative Areas ``Heavy Electrons''
(No. 20102008) of The Ministry of Education, Culture, Sports,
Science, and Technology, Japan.
The computation in this work has been done using the facilities
of the Supercomputer Center of Institute for Solid State Physics,
University of Tokyo.


\end{document}